\documentstyle[aps,epsfig,preprint,prl]{revtex}
\tightenlines

\begin{document}
\title{Action-derived molecular dynamics in the study of rare events }
\author{Daniele Passerone and Michele Parrinello}
\address{Max-Planck-Institut f\"ur Festk\"orperforschung, \\
Heisenbergstr. 1,
D-70569 Stuttgart, Germany}
\maketitle

\begin{abstract}
We present a practical method to generate classical trajectories with fixed
initial and final boundary conditions.
Our method is based on the minimization of a suitably defined
discretized action. The method finds its most natural application in the study
of rare events. Its capabilities are illustrated by non-trivial examples.
The algorithm lends itself to straightforward parallelization, 
 and when combined with {\it ab initio} 
molecular dynamics (MD) it promises to offer a powerful tool for the study of 
chemical reactions.
\end{abstract}


\vspace{0.5cm}

The dynamics of complex systems is often characterized by the occurrence of
rare but important events. Examples of these crucial events are chemical
reactions, diffusion processes and 
molecular conformational changes. The
reason for the slowness of these processes is often related to the
existence of large energetic barriers that the system has to clear in order
to pass from the initial to the final state.
Since this transition rate depends exponentially on the 
barrier height, the 
reaction characteristic time can exceed 
present-day computational capabilities by several orders of magnitude for 
barriers larger than $\simeq k_BT$.
 Most methods are based on the search
for the transition state starting from the initial configuration. The
dynamical properties are then reconstructed by making use of transition
state theory. In this class of methods one can include the traditional
quantum chemical approach \cite{chem}, conformational
flooding \cite{grub}, hyper-dynamics \cite{vot1} and
the parallel replica method \cite{vot2}. These
methods fail when transition state theory is not applicable. Furthermore
it often proves rather difficult to locate the transition state. This
is especially true in complex systems in which the potential energy surface
(PES) is very rugged and exhibits very many stationary points. In such a
case the very concept of transition state as a saddle point in the PES is
called into question. 

Elber and Karplus \cite{kaelb} have suggested an innovative approach.
Rather than starting from the initial state and looking for a transition state, they consider
a path that connects the initial and the final state. This approach 
is in spirit similar to the more modern nudged elastic band approach of
Jonsson \cite{jons}.
The method is appealing since it focuses on the global character
of the path rather
than on local properties of the PES. However, the paths 
do not reflect the real dynamics as the system crosses the
barrier. Following a seminal work of Pratt \cite{pratt}, the importance of determining the real dynamical path has been
emphasized very recently by Chandler and collaborators \cite{chand} who have
proposed a method for statistically sampling dynamical paths. This new
method is a very significant step towards the study of rare events in
complex systems, as it requires neither the assumption of transition
state theory, nor the existence of a single well defined transition state.
Its only prerequisite is the knowledge of at least one 
reactive dynamical trajectory. 
However, determining the initial trajectory is often difficult. A general strategy
suggested by Chandler and collaborators has been to anneal very unlikely trajectories. This procedure
is computationally demanding and does not guarantee that the system ends up in the desired final 
state.

Our work addresses this very general problem: given an initial and a final
configuration, what are the dynamical paths that connect them? In mathematical
terms this is a two-point boundary value problem. The standard boundary
conditions for Newton's equations fix instead the initial values of
positions and velocities. 

In principle the way to determine the dynamical path from configuration A to
configuration B has been known since 1744 when Maupertuis proposed the
principle of least action, published a few years later \cite{maup}. 
A more precise mathematical formulation is due to Hamilton.
In modern terms Hamilton's principle can be written as follows. 
Given a classical dynamical system
described by the set of coordinates ${\bf q}$, its trajectory ${\bf %
q}(t)$ with boundary conditions ${\bf q}(0)$= ${\bf q}_{A}$ and ${\bf q}%
(\tau)={\bf q}_{B}$ is determined by locating the stationary point of the
action:

\begin{equation}
S=\int_{0}^{\tau }L({\bf q}(t),\stackrel{\cdot }{{\bf q}}(t))\:dt,  \label{eq1}
\end{equation}
where $L$ is the Lagrangean $L=T-V$ and $T$ and $V$ are the kinetic and
potential energy respectively. By varying $S$ under the condition that the
initial and final points of the trajectory are fixed \cite{gold} Newton's
equations of motion follow. We want to turn this well-known principle into a
useful computational tool.
To this effect, following Gillilan and Wilson \cite{gillilan}
we discretize the trajectory into
P equispaced intervals and estimate the action as
\begin{equation}
S=\sum_{j=0}^{P-1}\Delta (\frac{1}{2}(\frac{{\bf q}_{j}-{\bf q}_{j+1}}{%
\Delta })^{2}-V({\bf q}_{j})),  \label{eq2}
\end{equation}
where ${\bf q}_{0}={\bf q}_{A}$ , ${\bf q}_{P}={\bf q}_{B}$ , $\Delta =\tau/P$
is the time interval in which we discretize $\tau $ and we are considering 
unitary masses.
Eq. (\ref{eq2}) is obtained by using the most primitive quadrature
for the action integral and by estimating the velocities from the difference
between successive points in the trajectory \cite{futurpassero}.

 The stationary point of this
action is expressed by a set of $P-1$ linear equations:

\begin{equation}
2{\bf q}_{j}-{\bf q}_{j-1}-{\bf q}_{j+1}-\Delta ^{2}\frac{\partial V({\bf q}%
_{j})}{\partial {\bf q}_{j}}=0,\label{eq3}
\end{equation}
that is, the stationary point of the discretized action is a Verlet trajectory,
 that is a trajectory that is identical to what would have been generated by the
well known Verlet algorithm. 
A straightforward search for the stationary point of $S$ (Eq. (\ref{eq2}))
is however very difficult, since the action is bounded neither from above nor
from below and we are looking for a stationary point, not necessarily a
maximum or a minimum. 
Furthermore, the nature of the stationary points depends on $\Delta$.
This makes the search for the solution of Eq. 
(\ref{eq3}) difficult, since it becomes a root-finding problem, which in order
to be solved requires a good guess of the solution and a Newton-Raphson 
procedure \cite{doll}.
The latter step involves the calculation of the second-order derivatives of 
the potential.

Alternatively, in order to find the solution of Eq. (\ref {eq3}) one
can look for the minimum of the function:

\begin{equation}\label{eqelber}
{\it O}=\sum_{j=1}^{P-1}(2{\bf q}_{j}-{\bf q}_{j-1}-{\bf q}_{j+1}-\Delta ^{2}\frac{\partial V({\bf q}%
_{j})}{\partial {\bf q}_{j}})^2,
\end{equation}
which is very close to the discretized version of the Onsager-Machlup (OM) action, as introduced by Olender and Elber \cite{elber}.
Locating the minimum of this action again involves 
the evaluation of second-order derivatives.
This, for an {\it ab-initio} molecular dynamics (MD) or for the study of
large systems within
ordinary MD, is a prohibitively expensive task.
Furthermore, in the practice the trajectories thus generated may
have a poor energy conservation,
especially in the region of the transition state, as we shall 
show in the following.

We present here a method which generates accurate
trajectories and requires only the calculation of
the forces. To this effect we shall use the fact
that the Newton trajectory has to conserve total energy.
This is also approximately true for the Verlet trajectories. If $\Delta $ is
sufficiently small the Verlet trajectories do not 
strictly conserve energy, but instead the energy oscillates around a 
roughly constant
value. We shall take advantage of this property
of the physically relevant trajectories and  
supplement the action with a penalty function which favors the energy
conserving trajectories:
\begin{equation}  \label{eq4}
\Theta({\bf q}_j,E)= S + \mu \sum ^{P}_{j=0}(E_{j}-E)^{2}.
\end{equation}
Here $\mu $ determines the strength with which we enforce energy
conservation, $E_{j}=({\bf q}_j-{\bf q}_{j+1})^2/(2 \Delta ^2)+V({\bf q}_j)$
is the instantaneous energy and $E$ its target
value.
Our definition of the kinetic energy is different from that which is normally
used in the Verlet algorithm, in which the velocities are estimated by the
more accurate central difference method. Nevertheless also $E_{j}$ is
approximately conserved and for simplicity's sake we shall use this definition
here. 
$E$ corresponds to the yet undetermined physical energy.

For large $\mu$ the second term in eq. (\ref{eq4}) dominates, and
the functional $\Theta({\bf q}_j,E)$ has a minimum.
For small $\mu$, on the other hand, $ S$ will determine the character of the
$\Theta$ extremum. The two regimes are separated by a critical value
$\mu^\star$.
In all the systems studied,
we have found that there exists a rather large interval 
of values of $\mu > \mu^\star$ such that $\Theta$ has a minimum and this 
minimum is globally very close to the Verlet trajectories.

Whenever the target energy $ E$ is not known, we can minimize $\Theta$
relative to
the ${\bf q}_j$ and to $E$. This procedure allows us to focus on the physically 
relevant values of $ E$.

In order to minimize $\Theta$ more efficiently, we follow
 \cite{doll} and make the transformation:

\begin{equation}\label{eqfou}
{\bf q}_j={\bf q}_{A}+\frac{j\,\Delta}{\tau }({\bf q}_{B}-{\bf q}%
_{A})+\sum_{n=1}^{P}{\bf a}_{n}\,\sin (\pi n\frac{j\,\Delta}{\tau}),
\label{fou}
\end{equation} 
thus automatically satisfying the boundary condition.
The advantage of using the ${\bf a}_n$ rather than the ${\bf q}_j$ is that the 
${\bf a}_n$ have a global character.
In practice,
we first optimize $\Theta$ with respect to a relatively small number
of ${\bf a}_n$, thus capturing the global feature of the trajectory, and then we add
the higher frequencies.
Each time we use a standard conjugate gradient algorithm to minimize $\Theta$.
This requires only the evaluation of the forces. The scaling is therefore
linear in the number of degrees of freedom rather than quadratic as 
in other approaches where second derivatives are needed.

We illustrate the performance of the functional $\Theta$
in a series of examples.
The most elementary one is a one-dimensional double well potential
already studied in ref. \cite{elber}.
In this simple case it is easy to find a Verlet trajectory oscillating
between the two minima.
If we start from this trajectory and add a random component, we
find that for $\mu=0$ a minimum or a maximum cannot be
found and the Verlet trajectory corresponds to a saddle point. This leads
to unstable behavior. We therefore
studied the behavior of $\Theta$ with respect to $\mu$.
We found that for $\mu<\mu^\star$
(which depends on $\Delta$) we could not locate a stationary point.
In particular, from the direct calculation of the Hessian matrix at the stationary point, we could extrapolate the change of sign of its lowest eigenvalue 
exactly at $\mu^\star$. The behavior of Euclidean distance in the close neighborhood of $\mu^\star$ foreshadows the closeness to the 
onset of the instability.

For larger values of $\mu$, however, we find solutions
very similar to the Verlet trajectory. In order to measure the 
accuracy of the $\Theta$ solution, we evaluated the norm of the 
distance 
(in the $(P-1)-$dimensional space of the trajectories)
between the Verlet trajectory and the minimum, as well as the function
$O$ in eq. (\ref{eqelber}).
The results shown in Fig. 1 give a comparable but not
identical estimate for the optimal
value of $\mu$. It must be stressed, however, that to all 
interests and purposes, the variational solutions and the original
Verlet trajectory are globally very close, which makes the 
precise choice of $\mu$ not critical.
This is very important, since in most cases the exact trajectory is not 
known;
in such cases only the criterion of minimal $O$ can be used to optimize $\mu$.

The second example is a minimization of a trajectory in a two-dimensional configurational 
space, namely in the Mueller potential \cite{mueller}, which was invented as a 
non-trivial test for reaction path algorithms: it has two main minima, and a
third minimum that somewhat hinders the
trajectory from seeing the saddle point. 
The aim of the optimization is to find a physical
path between the two deepest minima, possibly locating the saddle point.

We directly compared our results with those of ref. \cite{elber}. Both 
sets of trajectories lead to an apparently satisfactory result. 
They pass exactly through the saddle point,
and the overall behavior of the trajectories is clearly physical.
Our trajectory has of course a larger value of $O$. 
However, the $O$-trajectory shows poorer energy 
conservation close to the transition state, as we see in Fig. 2.
This is due to the fact that the optimization of $O$ is global and 
local errors leading to energy non-conservation have little weight. Of course
one could also supplement $O$ with a penalty function to improve the 
energy conservation; this would still require the calculation of 
second-order derivatives.

As a last and far less trivial example we look at
a process in which the central atom of a seven-atom two-dimensional
Lennard-Jones cluster migrates to the surface. This process has been
studied in detail by Dellago {\it et al.}  \cite{chand2}. Our calculations
reproduce their highly non-trivial results:
in Fig. 3 we show how two of the paths we found correspond to the most probable
ones found through 
the elaborate procedure of 
path sampling in \cite{chand2}. As can be seen in
Figs. 3 and 4, the dynamical path has to pass several transition states.
This is  a severe challenge for  other methods.

It has been suggested by Zaloj and Elber \cite {elber2}
that the $O$-trajectories can be used 
to increase the
value $\Delta$ beyond what is normal in ordinary MD.
The basic assumption is that by so doing,
the effect of the higher frequencies is integrated out and leads to a 
Gaussian distribution of the errors:

\begin{equation}\label{eqeps}
\epsilon_j= 2{\bf q}_{j}-{\bf q}_{j-1}-{\bf q}_{j+1}-\Delta ^{2}\frac{\partial V({\bf q}%
_{j})}{\partial {\bf q}_{j}}.
\end{equation}
With this in mind, we plot in Fig. 4
the distribution of the errors $\epsilon_j$, defined by (\ref{eqeps}). 
As is exemplified by the figure, we verified that in the $\Theta$ trajectories,
the $\epsilon_j$ appear to be normally distributed. Thus
one can implement an accelerated MD scheme in the spirit of ref. \cite{elber2}.

In conclusion, we have presented a novel method for generating
dynamical trajectories with pre-assigned initial and final boundary
conditions. The method can be easily implemented within the Car-Parrinello MD
approach,
offering a powerful tool for the study of chemical reactions, and
can be combined with path sampling.
Its advantages over previous schemes are the fact that
it requires only the calculation of the forces, 
 its numerical stability and the quality of 
the trajectories: the latter 
is a direct product of  the energy conservation constraint.

As a side effect, the method can lead to easy localization
of the saddle points without losing the physical soundness of the solutions.
Our approach lends itself to very
efficient parallelization.
Multiple time-scale approaches are natural and feasible.
We are sure that many other applications of
our method will be found and we believe that it is a noteworthy advance
in the field of molecular dynamics.

\begin{sloppypar}
It is a pleasure
to thank W. Silvestri, A. Gambirasio, M. Bernasconi, and F. Filippone 
for fruitful discussions and suggestions.
The research
of D. P. at MPI is sponsored by the Alexander von Humboldt Foundation.
\end{sloppypar}

\begin{figure}[tbp]
\caption{Euclidean distance between the $\Theta$ trajectory and the practically 
exact Verlet trajectory (red, left scale) and OM action value (blue, right 
scale) as a function of $\mu$ in a one-dimensional double-well potential of the 
type 
$U(x)=1/2(x^2+A*\exp(-\alpha x^2)-B)$, with $B=1/\alpha[\log(\alpha A)+1]$,
$A=80$ and $\alpha=0.04375$ {\protect \cite {elber}}.
With these parameter values,
the two minima are located at $\pm 5.33$ and the barrier height is $14.2$.
The integration time step is $\Delta=0.31$ and the total number of $P$ points
is 64. The starting guess for the minimization of $\Theta$ is a randomized 
Verlet trajectory. For $\mu=\mu^{\star}=0.39$ the Hessian of $\Theta$ ceases
to be positive definite. In the inset, the Verlet trajectory (red) is 
compared with our solution (blue) at $\mu=100$.
}
\end{figure}
\begin{figure}[tbp]
\caption{Upper panel: $\Theta$ trajectories (blue crosses) and O-trajectories
(red dots) on the Mueller potential energy surface {\protect \cite{elber}}.
The potential energies of the two lowest
minima are -146 and -106, and the transition 
state is located at -41 (same units as in {\protect \cite{elber}}).
Lower panel: energy as a function of time for our trajectory (blue line) and
O-trajectory (red line). As in reference {\protect \cite{elber}} we have taken
a time step of 0.01; the total number of points was 300.
The regions of accumulation of blue points correspond to the two lowest minima.
}
\end{figure}
\begin{figure}[tbp]
\caption{Two of the paths followed by a two-dimensional cluster of 7 atoms
for the migration of a central atom to the surface. In Lennard-Jones units
the total time is 3 and the time step was taken to be $\Delta=0.06$. The 
initial trajectory was a linear interpolation between the initial and final 
points. We show in the picture the sequence of metastable states visited
by the system.
}
\end{figure}
\begin{figure}[tbp]
\caption{Potential energy profile (top) and conserved energy (middle)
for the path $(b)$ in Fig. 3.
The two intermediate metastable states depicted in Fig. 3
are indicated by arrows. The lower panel shows the distribution of the errors
$\epsilon_j$ along the same trajectory. The green line shows a gaussian fit.
}
\end{figure}

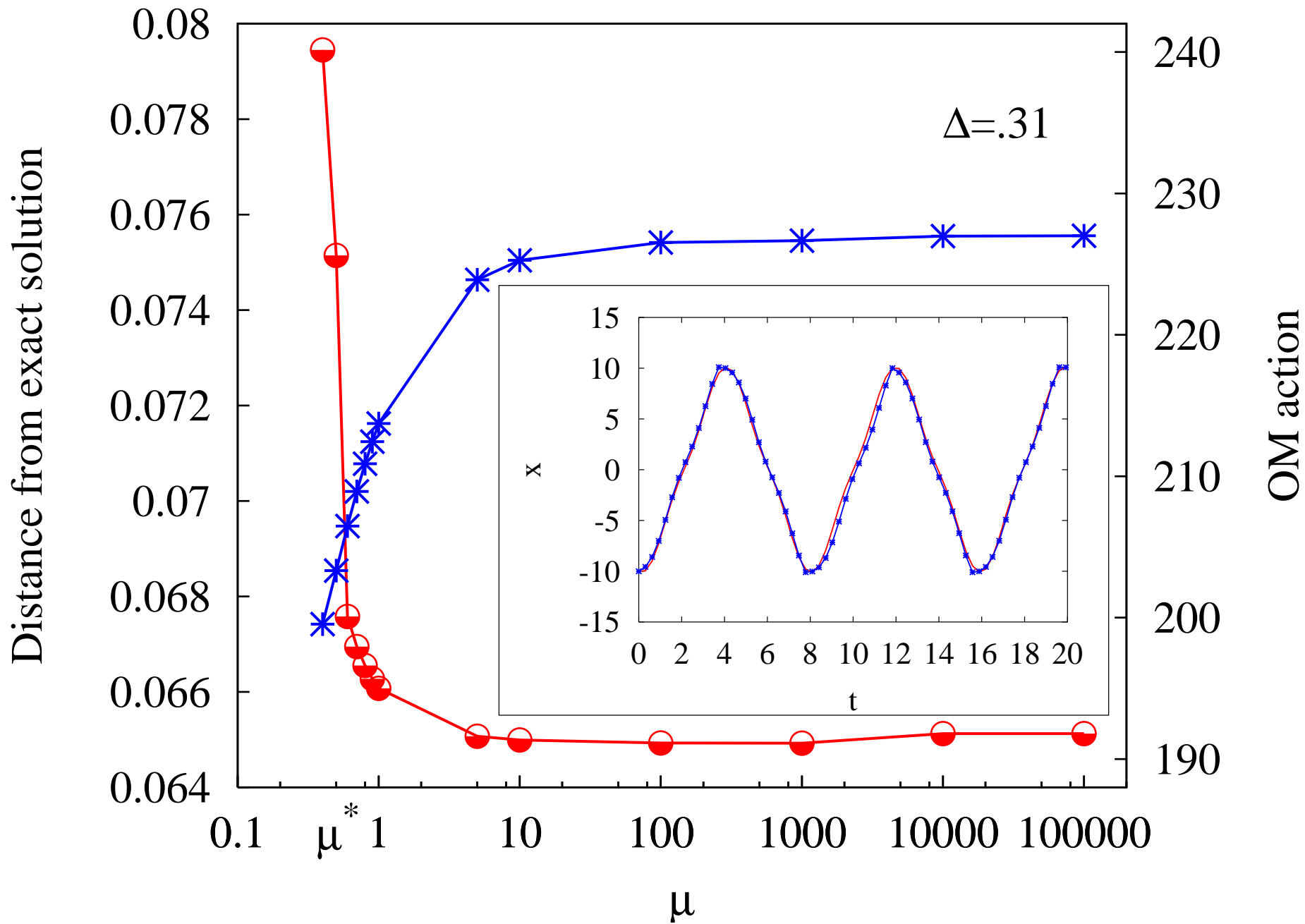
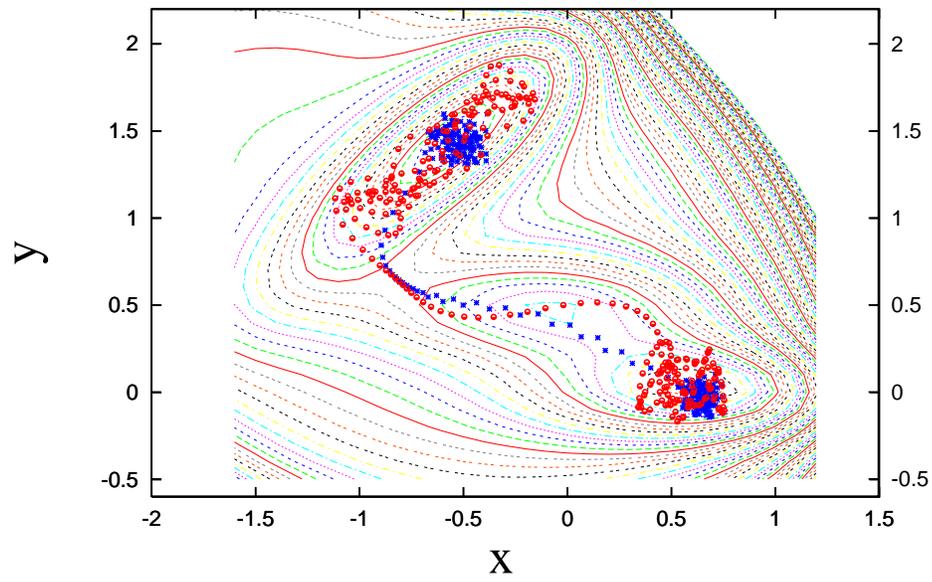
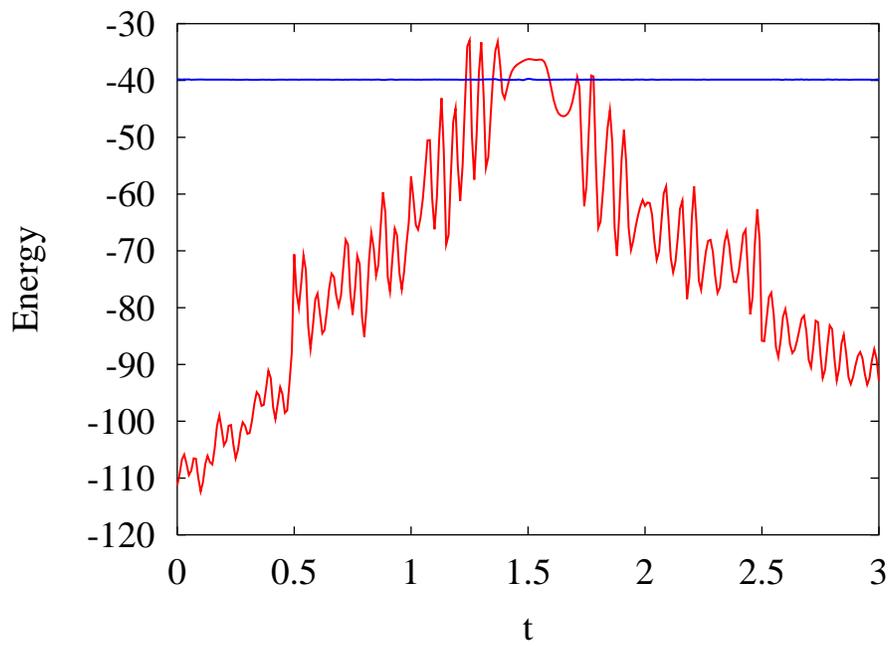
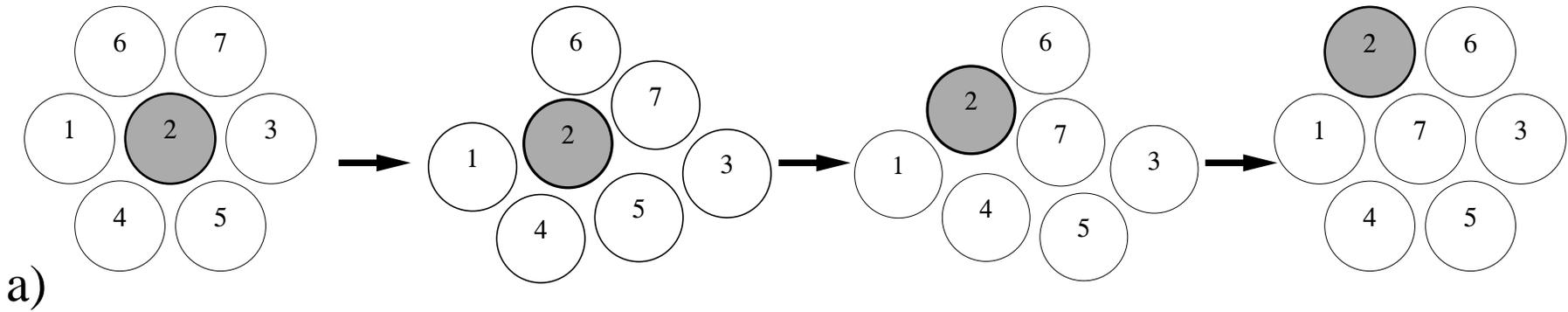
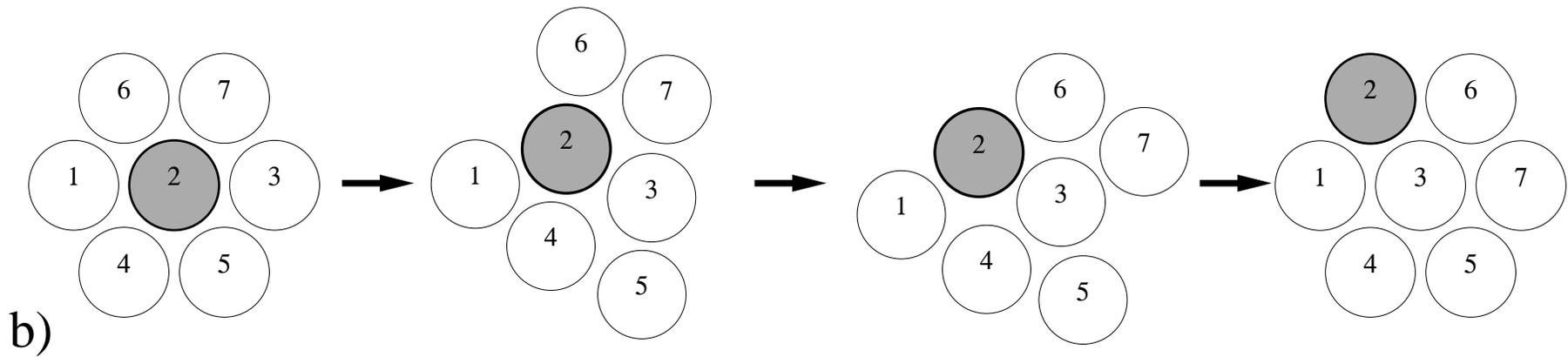
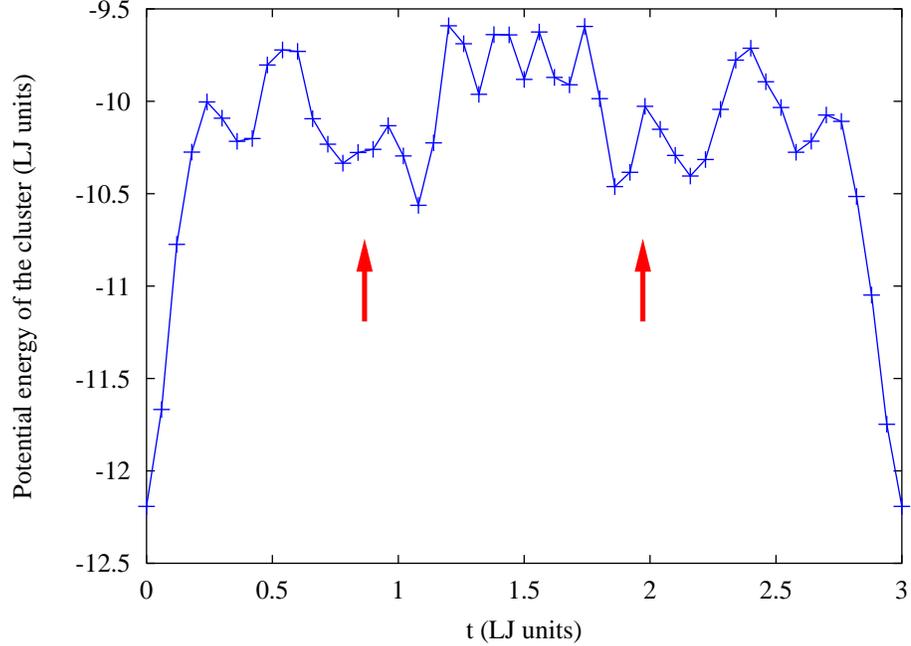
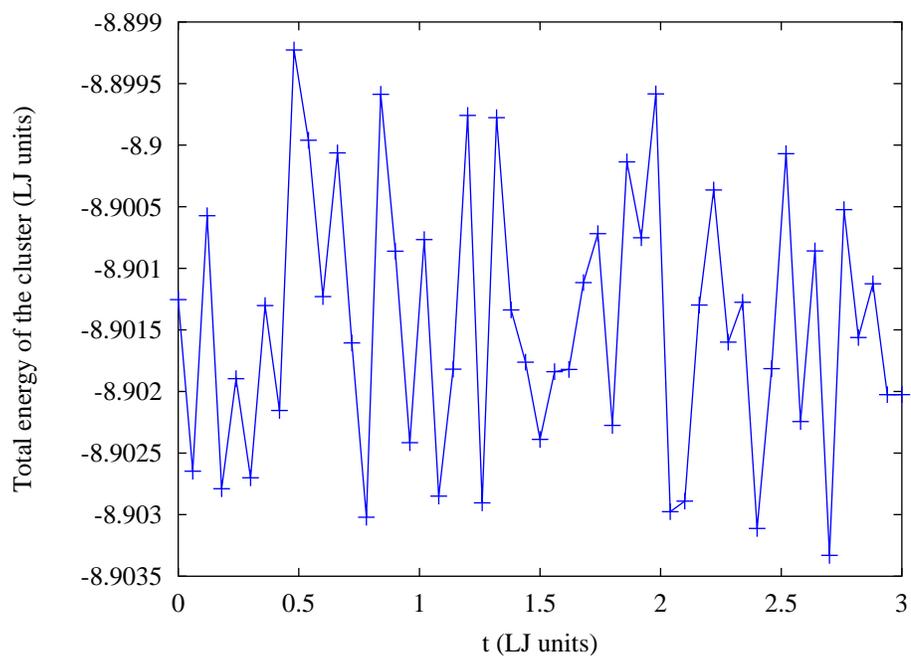
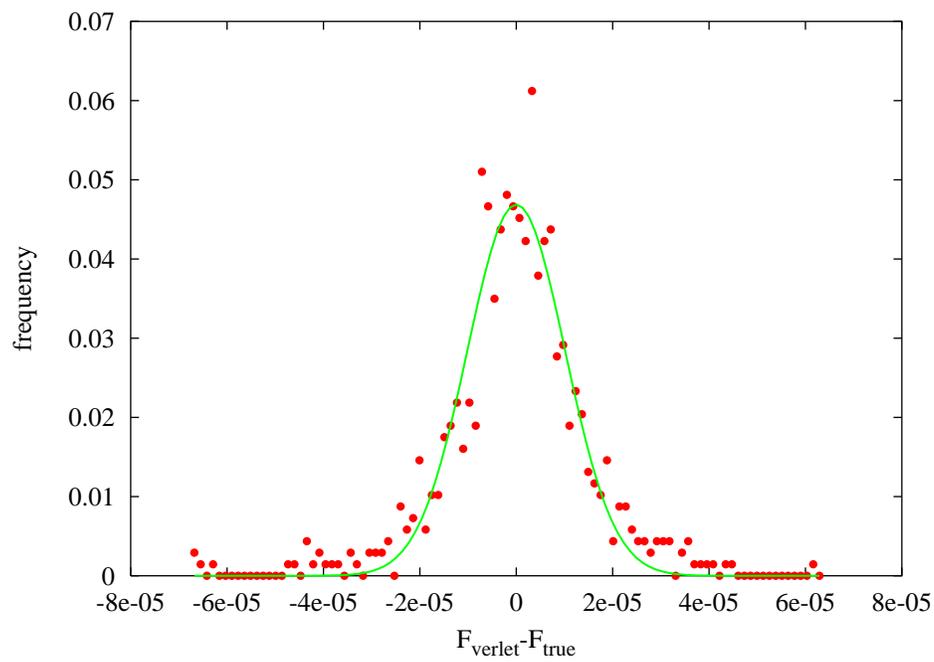
\end{document}